\documentclass[letterpaper, 10 pt, conference]{ieeeconf}  

\IEEEoverridecommandlockouts                              

\usepackage{footmisc}
\usepackage{graphics} 
\usepackage{epsfig} 
\usepackage{amsmath} 
\usepackage{amssymb}  
\usepackage{bm}
\usepackage{graphicx}
\title{\LARGE \bf
Torque Controlled Locomotion of a Biped Robot with Link Flexibility
}

\author{Nahuel A. Villa$^1$, Pierre Fernbach$^2$, Maximilien Naveau$^{1,2}$, Guilhem Saurel$^1$,\\ Ewen Dantec$^1$, Nicolas Mansard$^{1,3}$, Olivier Stasse$^{1,3}$
\thanks{* For this work N. A. Villa, O. Stasse were supported by the cooperation agreement ROB4FAM. M. Naveau, G. Saurel and N. Mansard were supported by the H2020 Memmo project. P. Fernbach by the cooperation agreement DynamoGrade, E. Dantec was supporter by ANITI.}
\thanks{$1$ Gepetto Team, LAAS-CNRS, Université de Toulouse, France.}
\thanks{$2$ TOWARD, Toulouse, France.}
\thanks{$3$ Artificial and Natural Intelligence Toulouse Institute, France.}%
\thanks{e-mails: \href{mailto:nahuel.villa@laas.fr}{nahuel.villa@laas.fr},\;\; \href{mailto:pierre.fernbach@toward.fr}{pierre.fernbach@toward.fr},}%
\thanks{\hspace{10mm}\texttt{firstName.lastName@laas.fr}}
}

\usepackage{url}

\makeatletter
\let\NAT@parse\undefined
\makeatother
\usepackage{hyperref}
\hypersetup{
	colorlinks=true,
	linkcolor=[rgb]{0.015, 0.59, 0.77},
	citecolor=[rgb]{0.78, 0.03, 0},  
	urlcolor=[rgb]{1, 0.612, 0.03}
}

\begin{document}

\maketitle
\thispagestyle{empty}
\pagestyle{empty}

\begin{abstract}

When a big and heavy robot moves, it exerts large forces on the environment and on its own structure, its angular momentum can vary
substantially, and even the robot's structure can deform if there is a mechanical weakness. Under these conditions, standard locomotion controllers can fail easily. 
In this article, we propose a complete control scheme to work with heavy robots in torque control. The full centroidal dynamics is used to generate walking gaits online, link deflections are taken into account to estimate the robot posture and all postural instructions are designed to avoid conflicting with each other, improving balance. These choices reduce model and control errors, allowing our centroidal stabilizer to compensate for the remaining residual errors. The stabilizer and motion generator are designed together to ensure feasibility under the assumption of bounded errors.
We deploy this scheme to control the locomotion of the humanoid robot Talos, whose hip links flex when walking.
It allows us to reach steps of 35~cm, for an average speed of 25~cm/sec, which is among the best performances so far for torque-controlled electric robots.

\end{abstract}

\section{INTRODUCTION}

Legged robots are normally modeled and controlled as a chain of rigid bodies with actuated joints connecting them \cite{Handbook}. This simplification of the structural material properties is specially accurate to deal with robots that are light or have multiple legs \cite{Leziart2021_Solo12}. Nevertheless, heavy biped robots such as Talos or Walkman ($\approx 100$~kg) can present small but meaningful deflections of their structure. These unmodeled deflections produce a bad estimation of contact points as well as a slow transference of forces through the kinematic chain, resulting, therefore, in wrong contact forces and a bad tracking of the desired robot motion. Due to the unstable dynamics of legged robots, the tracking error tends to grow, ending up with a control failure.

Flexible components are the subject of several studies in robotics in general \cite{DeLucca2016Flexibility} and humanoids in particular:

Flexible joints based on Series Elastic Actuators (SEA) \cite{Pratt2002SEAs, PetitDA15} have been used and studied on humanoid robots such as Walkman \cite{negrello:icra:2016}, Coman \cite{Lee2013COMAN_SEAS} or Valkirie \cite{Paine2015ValkyrieActuation} which, thanks to joint sensors, take advantage of the flexibility for safe environment interaction, disturbance rejection and dissipation of walking impact energy. In our case, however, deflections are not directly measurable as they are produced on the robot links.

Flexible bodies use to be incorporated to the end effectors of position-controlled robots to measure interaction forces from their deflection and to damp walking impacts, such as in the HRP series \cite{Benallegue2015_HRP-2_flexEstim, Caron2019stair, Napoleon2005analysis}. The locomotion of these robots is normally controlled with approaches derived from \cite{kajita:icra:2001}, where the deflection is estimated based on the desired contact forces. 

\begin{figure}[t!]
\centering
\includegraphics[width=.45\textwidth]{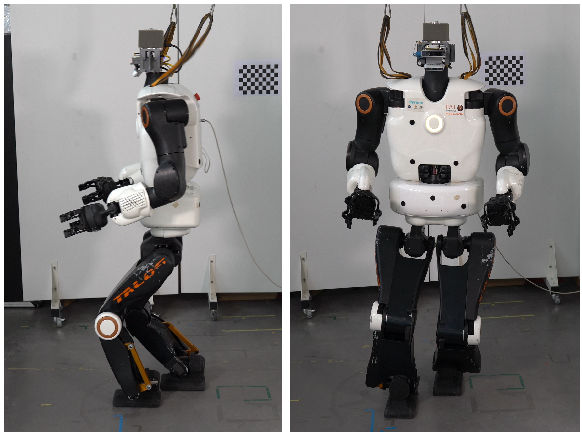}
\caption{\!\!\!\!\href{https://gepettoweb.laas.fr/articles/talos_centroidal_mpc_torque_control.html}{Snapshots} of Talos walking dynamically, in torque-control, with a calmed velocity of $15$~cm/s.}
\label{Fig.Walking}
\end{figure}


Similar to the later case, in this article, we estimate the link deflections based on the commanded joint torques to avoid the noise and delay introduced with the measurement of torques. Using a rigid-robot model, we incorporate such deflections in the closest joints to obtain a better posture estimation. We also reduce typical approximation errors by previewing all centroidal non-linear behaviors in our motion planning scheme. We obtain even further reduction of the control errors by making all references of the inverse dynamics consistent with the full centroidal motion and with each other.

The remaining (much smaller) model error, as well as all internal and external disturbances, produce tracking errors that grow with the robot dynamics. We use state feedback to stabilize the behavior of the Center of Mass (CoM) of the robot and, based on a reachability analysis of the resulting closed-loop system \cite{Villa2019Sensitivity2Uncertainties}, we deploy a tube-based MPC \cite{Tube} scheme that guarantees robust feasibility when disturbances are bounded.

In particular, we use the robot Talos, shown in Fig.~\ref{Fig.Walking}, as an experimental platform for this work. Talos is a commercial humanoid robot equipped with powerful actuators and precise sensors in a strong structure \cite{Stasse2017Talos}. Nevertheless, this platform has shown high difficulties to develop repeatable walking gaits in torque control. As an outcome of this work, we can claim that such difficulties are largely explained by the unforeseen flexibility that the model Talos exhibit on its hip links. 

We propose a simple method to identify the stiffness on the robot hips and using the identified values on the proposed control scheme, we have obtained fast and dynamic walking gaits on torque control with velocities up to $25$~cm/s, the fastest walking velocity reached on Talos up to date.


Following this introduction, Section~\ref{Sec.Model} describes the robot dynamics and flexibility. Section~\ref{Sec.CompensationOfDeflections} details our approach to incorporate deflections in the posture estimation. Our centroidal controller is exposed in Section~\ref{Sec.BalanceControl}. The stable centroidal motion is mapped into joint torques by the whole-body controller, as explained in Section \ref{Sec.WholeBodyControl}. We deploy all the previous control infrastructure on the robot Talos to perform the experiments discussed in Section~\ref{Sec.Experiments}, and Section \ref{Sec.Conclu} summarizes our main conclusions.

\section{MODELING}\label{Sec.Model}
\subsection{Whole-body dynamics}
Walking robots are normally represented as a kinematic chain of $ n $ joints connecting $ n\texttt{+}1 $ links, in which no link is attached to the inertial world frame \cite{Nava2016Stability}. The robot configuration $ q\!\!~=~\!\![\begin{matrix} 
q_{w}^{\top}& q_{j}^{\top}
\end{matrix}]^{\top}$ can be described by the position and orientation of the base link (robot waist) $ q_{w}\!\!\!~\in~\!\!\!\mathit{SE}(3)$, and the posture given by all joint angles $ q_{j}\in \mathbb{R}^{n} $.

Joint motors produce the torques $ \tau_{a}\in \mathbb{R}^{n} $, required for the robot motion, following the dynamics \cite{Handbook}:
\begin{equation}\label{eq.LagrangeanDynamics}
M(q)\ddot{q} + h(q,\, \dot{q}) = S_{j}\tau_{a} + \sum_{k} J_{k}(q)^\top f_{k},
\end{equation}
where $ M(q) \in \mathbb{R}^{(n\texttt{+}6)\times(n\texttt{+}6)} $ is the generalized inertia matrix, $ h(q, \dot{q}) \!\!\in \mathbb{R}^{n\texttt{+}6} $ stands for Coriolis, centrifugal and gravity forces, $ S_j $ selects all directly actuated joints, and for the $k$-th contact, $f_k\in\mathbb{R}^3$ is a force exerted by the environment on the point associated to the Jacobian matrix $J(q)^{\top}_{k}\!\!\in\mathbb{R}^{n\texttt{+}6\times 3}$.

Joint angles must lie on collision-free ranges, and joint torques are limited by the employed motors and materials:
\begin{align}
q^{\mathrm{min}}_{j}\leq q_{j}^{ }&\leq q^{\mathrm{max}}_{j},\label{eq.jointLimits}\\
\tau^{\mathrm{min}}_{a}\leq\tau_{a}&\leq\tau^{\mathrm{max}}_{a},\label{eq.torqueLimits}
\end{align}
where the inequalities with lower and upper limit vectors hold element-wise.

We assume that feet $s$ do not slide during ground contacts:
\begin{align}\label{eq.StaticContacts}
\dot{s} &= J_{s}\dot{q},\\
\ddot{s} &= J_{s}\ddot{q} + \dot{J}_{s}\dot{q}=0, \label{eq. NonSlidingCondition}
\end{align}
and that ground contact forces are unilateral, constrained to friction cones of the form \cite{Handbook}:
\begin{equation}\label{eq.Unilaterality and Friction}
\Vert f_{k}^{p}\Vert \leq \mu f_{k}^{n}\quad \forall f_k ~\text{ in the ground-feet contact},
\end{equation}
where the friction forces $f_{k}^{p}$ parallel to the contact surface are limited by the normal force $f_{k}^{n}$ with some friction coefficient $\mu > 0$.

\subsection{Centroidal dynamics}
Balance and locomotion dynamics corresponds to the under actuated part of~\eqref{eq.LagrangeanDynamics} and can be isolated from the posture dynamics without additional assumptions~\cite{Wieber2006holonomy,herzog2015trajectory}. Let us consider a Cartesian coordinate system with the origin on some ground contact surface, and the axis $ ^z $ aligned with the gravity. So, in the lateral coordinates $^{xy}$, this dynamics relates the motion of the Center of Mass (CoM) $c$ of the robot to the Center of Pressure (CoP)~$p$ of the ground contact forces \cite[Chapter 2]{Villa2019managing} as
\begin{equation}\label{eq.CentroidalDynamics}
    p^{xy} = c^{xy} - \dfrac{mc^{z}\ddot{c}^{xy}-S\dot{L}^{xy}}{m(\ddot{c}^{z}+g^{z})} + \dfrac{\sum_k r_{k}^{z} f_{k}^{xy}}{\sum_k f_{k}^{z}},
\end{equation}
where $g$ is the vertical acceleration due to gravity, $m$ is the total robot mass, $\dot{L}$ is the variation of the angular momentum, $r_k$ are the ground contact points and $ S=\big[\begin{smallmatrix}\,0\, & \texttt{-}1\\1\, &0
\end{smallmatrix}\big]$ is a $\frac{\pi}{2}$ rotation matrix. Due to unilaterality of ground contact forces \eqref{eq.Unilaterality and Friction}, the CoP is bound to the support polygon $\mathcal{P}$ \cite{Handbook}:
\begin{equation}\label{eq.DynamicConstraint}
    p\in\mathcal{P}(s)
\end{equation}
that varies depending on the current foot positions $s$.

\subsection{Hip flexibility}

Deflections on Talos' hips are concentrated on its waist-leg connection, where the link cross-section narrows. This introduces extra degrees of freedom that can be represented with passive virtual joints \cite{Nakaoka2007constraint}. As deformations appear on vertical linkages, we only model pitch and roll deflections, which produce the main impact on foot placement. We obtain in this form a model for the robot Talos with 42 degrees of freedom composed by 32 actuated joints, 4 elastic passive joints and the global position and orientation of the robot.

We built a simulator for this model using \href{https://github.com/stack-of-tasks/pinocchio}{\texttt{pinocchio}} \cite{Pinocchioweb} for the computation of rigid body dynamics and the \href{https://github.com/Gepetto/gepetto-viewer-corba}{\texttt{gepetto-viewer}} for visualizations, as in the companion \href{https://gepettoweb.laas.fr/articles/talos_centroidal_mpc_torque_control.html}{video}\footnote{video available in: \url{ https://gepettoweb.laas.fr/articles/talos_centroidal_mpc_torque_control.html}\label{video}}, where we have included massless collision-free plates on the virtual joints to display deflections.

We simulate the elastic deformation of virtual joints as a spring damper 
\begin{equation}
    \tau_f^{ } = -k_f \theta - d_f \dot{\theta}, \label{eq.ElasticBehaviour}
\end{equation}
with stiffness $ k_f $ and damping $ d_f $ coefficients that relate deflections $\theta$ to the flexing torque $ \tau_f^{ } $. As the identified values of stiffness (see Section~\ref{Sec.stiffnessIdentification}) are relatively big, simulations require short integration periods ($ \approx 0.1 $~ms) for numerical convergence. 


\section{POSTURE ESTIMATION WITH DEFLECTIONS}\label{Sec.CompensationOfDeflections}

\begin{figure}
	\centering
	\includegraphics[trim={0 1mm 0 0},clip,width=.44\textwidth]{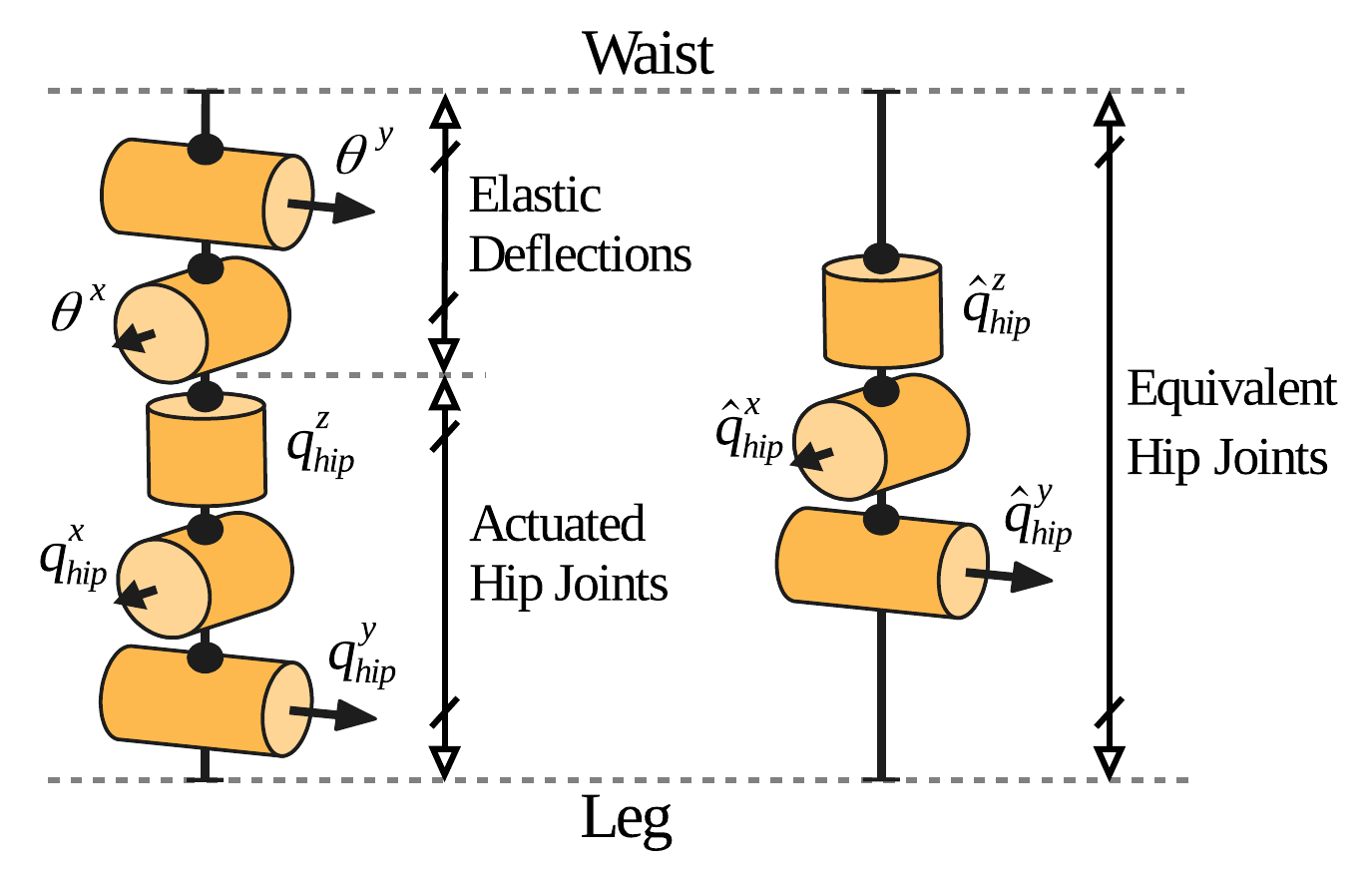}
	\caption{\textbf{Talos hips:} We estimate the robot posture by condensing the 5 hip rotations observed on the flexible robot into 3 rigid-model joints that produce the same total rotation.} \label{Fig.HipJoints}
\end{figure}

Hip configurations, outlined in Fig.~\ref{Fig.HipJoints}, are composed by 3 measured joint rotations $ q_\mathit{hip}^{ } \in \mathbb{R}^3 $ and 2 unmeasurable elastic deflections $\theta \in \mathbb{R}^2 $ that can be estimated.
We condense all this chain into an equivalent hip configuration\footnote{Code available on GitHub: \href{ https://github.com/Gepetto/flex-joints}{\texttt{Gepetto/flex-joints}}.} $\hat{q}_\mathit{hip}^{ } \in \mathbb{R}^3 $ with only $3$ actuated joints that satisfy
\begin{align}
    &R(\hat{q}_\mathit{hip}^{zxy}) = R(\theta^{yx})R(q_\mathit{hip}^{zxy}),\\
    &\hat{\dot{q}}_\mathit{hip} = \Big(S^{z}+R(\hat{q}_\mathit{hip}^{z})S^{x}+R(\hat{q}_\mathit{hip}^{zx})S^{y} \Big)^{\!\texttt{-}1}\hat{\omega}_\mathit{hip},
\end{align}
where the selection matrix $S^i \in \mathbb{R}^{3\times 3}$ is null except for its $i$-th diagonal element that is 1, $ R(\cdot) \in \mathbb{R}^{3\times 3}$ is a rotation matrix, and the equivalent angular velocity is 
\begin{equation}
    \hat{\omega}_\mathit{hip} = S^{y}\dot{\theta} + R(\theta^{y})S^x\dot{\theta} + R(\theta^{yx})\omega_\mathit{hip}.
\end{equation}

We estimate hip deflections $\theta$ from \eqref{eq.ElasticBehaviour} as  
\begin{align}\label{eq. Deflections}
    \theta &= \dfrac{d_f\theta_0 - \tau_f^{ }\,\Delta t}{k_f\,\Delta t+d_f},& & \dot{\theta} = \mathrm{LPF}\bigg(\dfrac{\theta - \theta_0}{\Delta t}\bigg),
\end{align}
where the rate of change of deflections is Low Pass Filtered (LPF) to avoid big jumps coming from the numerical derivative, $ \theta_0 $ is the previous-estimated deflection, $\Delta t$ is the posture sampling period, and we approximate the flexing torque $\tau_f^{ }\in \mathbb{R}^2$ with the commanded hip torque $\tau_{hip}^{ }$ and expected hip force $f_\mathit{hip}$:
\begin{equation}
    \tau_{f} = \left[R(q_\mathit{hip}^z)S^x\tau_\mathit{hip} + R(q_{hip}^{zx})S^y\tau_\mathit{hip} + R(\theta^{yx})l\times f_\mathit{hip}\right]^{xy},
\end{equation}
considering the distance $l$ between hip joint and concentrated deflection. In the case of Talos, $l = [\begin{smallmatrix} 0&0&9 \end{smallmatrix}]$~cm for both hips.
This simple approximation of torque and deflections can be improved as we propose in \cite{Giulio2022_TalosFlex}.

Notice that the resulting estimated posture has still inaccuracies due to the approximated flexing torque and the reduced degree of freedom of the equivalent rigid posture (\href{ https://gepettoweb.laas.fr/articles/talos_centroidal_mpc_torque_control.html}{video}\footref{video}), but such uncertainty is small enough for the centroidal stabilizer (described in the following) to deal with it. 

\section{BALANCE AND LOCOMOTION CONTROL}\label{Sec.BalanceControl}

\begin{figure*}[th]
	\centering
	\includegraphics[trim={0 1mm 0 3mm},clip,width=.8\textwidth]{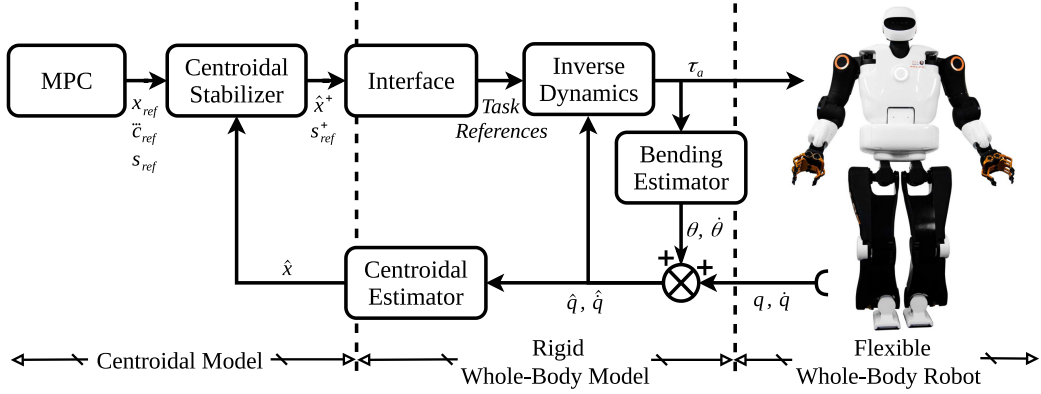}
	\caption{\textbf{Control Diagram:} It is separated in two abstraction levels, the centroidal controller, on the \textit{left}, generates and stabilizes the CoM trajectory. \\\hspace*{33mm}The whole-body controller, on the \textit{right}, generates accordingly all joint torques required for the desired locomotion.} \label{Fig.ControlDiagram}
\end{figure*}

The general organization of our control scheme is given in Fig.~\ref{Fig.ControlDiagram}.
In this section, we focus on the centroidal control and stabilization.

\subsection{Linearized centroidal dynamics}

We rewrite the centroidal dynamics with the linear form:
\begin{equation}\label{eq.LinearizedCentroidalDynamics}
\ddot{c}^{xy} = \omega^{2}(c^{xy}-v^{xy}),
\end{equation}
with some constant value $\omega^{2}\approx{g^{z}}\!/{c^{z}}$, by introducing the compensated CoP \cite{Villa2019managing}, also called Virtual Repellent Point (VRP)~\cite{Englsberger2015_3D} $v$, which requires us to estimate the bias term 
\begin{align}
    n^{xy} &\triangleq p^{xy}-v^{xy} \label{eq.VRP_and_COP} \\ &= \dfrac{\ddot{c}^{xy}}{\omega^{2}}  - \dfrac{mc^{z}{\ddot{c}}^{xy} - {S}{\dot{L}}^{xy}}{m(\ddot{c}^{z}+ g^{z})} + \dfrac{\sum_k r_{k}^{z} f_{k}^{xy}}{\sum_k f_{k}^{z}} , \label{eq.Non-linearEffects}
\end{align}
based on the previous motion. When the robot is on a flat and horizontal ground, this bias can be obtained out of the recursive Newton-Euler algorithm as proposed in \cite{naveau:metapod:2014}%
, or from the centroidal momentum matrix \cite{Orin2008Centroidal}.

As this dynamics is identical and decoupled in $^{x}$~and~$^{y}$ coordinates, we drop the superscript while keeping the following valid for any $^x$ or $^y$.
Then, from \eqref{eq.LinearizedCentroidalDynamics}, we obtain a linear time-invariant sampled-time control system
\begin{equation}\label{eq.SystemDynamics}
x^{+} = Ax + Bu,
\end{equation}
with the state and system matrices:
\begin{equation}
x = \!\!\begin{bmatrix}
c\\\dot{c}\\ \ddot{c}
\end{bmatrix}\!\!,\;\;\;
A =\!\! \begin{bmatrix}
1&T&\frac{T^2}{2} \\
0&1&T \\
0&0&1
\end{bmatrix}\!\!,\;\;\;
B =\!\! \begin{bmatrix}
\frac{T^3}{6}\\
\frac{T^2}{2}\\
T
\end{bmatrix}\!\!, \label{eq.StateConstraint}
\end{equation}
by commanding piecewise constant jerks $ u = \dddot{c} $ over time intervals $T$ \cite[Chapter 5.5]{Ogata}. Here, the VRP is a system output $v=Vx$ with $V=\left[\begin{smallmatrix}1& 0 & \texttt{-}\omega^{\texttt{-}2} \end{smallmatrix}\right]$, and feasible states (denoted $x\in\mathcal{X}$) must satisfy the unilaterality of forces \eqref{eq.DynamicConstraint}:
\begin{equation}
    Vx\in \mathcal{P}(s) - n. \label{eq.DynamicalConstraint}
\end{equation}

\subsection{Reference motion}

The robot is controlled to follow some feasible trajectory of states that can be predefined off-line \cite{Choi, Song2015_CharacterReference}, or online, typically with an MPC approach \cite{Mayne2000constrained, Herdt2010online, Wieber2006}.
In this article, we simply deploy the MPC scheme proposed in \cite{Herdt2010online}, constrained to satisfy the system dynamics \eqref{eq.SystemDynamics}
\begin{equation}\label{eq.ReferenceTrajectory}
    x_\mathit{ref}^{+} = Ax_\mathit{ref} + Bu_\mathit{ref}, 
\end{equation}
with feasible states
\begin{equation}\label{eq. ReferenceStateConstraint}
    x_\mathit{ref} \in \mathcal{X} \ominus \Omega,
\end{equation}
where $\ominus$ is a Pontriagyn difference\footnote{Given sets $A$ and $B$, $\;\;A\ominus B \triangleq \lbrace x \,|\; x + B \subseteq A\rbrace$.}, used to introduce the safety margin $\Omega\subset\mathcal{X}$ (see Sec. \ref{Sec.Stabilizer}).
The step placements are kinematically constrained by the leg lengths in some stepping area
\begin{equation}
    s_\mathit{ref} \in \mathcal{S},
\end{equation}
with limited swinging foot velocity $\dot{s}_\mathit{ref} \leq \dot{s}_\mathrm{max}$. Moreover, in order to maintain stability and recursive feasibility of the planned motion, we restrict the terminal state $x_\mathit{ref}^\mathit{term}$ to satisfy the capturability condition \cite{SherikovThese}:
\begin{equation}
    \xi_\mathit{ref}^\mathit{term}\in \mathcal{P}(s^\mathit{term}_\mathit{ref})\ominus V\Omega - n^\mathit{term},
\end{equation}
where $V\Omega$ is the CoP safety margin\footnote{Given the vector $a\in\mathbb{R}^n$ and set $B\subset\mathbb{R}^n$, $\;\;aB \triangleq \lbrace a x \,|\; x \in B\rbrace$.}, and $\xi =  c + \omega^{\texttt{-}1}\dot{c} $ is called divergent component of the motion \cite{Handbook}.

Within these restrictions, the planned motion is optimized to track some aimed CoM velocity $ \dot{c}_\mathit{ref}\! \rightarrow\! \dot{c}_\mathit{aim} $ with minimum jerk $\dddot{c}_\mathit{\!ref} \rightarrow 0 $ and ankle torque $v_\mathit{ref}\rightarrow  \texttt{-} n$. 

\subsection{Centroidal Stabilizer}\label{Sec.Stabilizer}

We track the reference motion $ x_\mathit{ref} $, $ u_\mathit{ref} $ \eqref{eq.ReferenceTrajectory} by commanding the desired next state $\hat{x}^+$ (see Fig.~\ref{Fig.ControlDiagram}) computed as:
\begin{equation}
    \hat{x}^{+} = A\hat{x} +Bu_\mathit{ref} + B\,\mathrm{sat}\big(K(\hat{x} - x_\mathit{ref})\big),
\end{equation}
with a saturated linear feedback $\mathrm{sat}\big(K(\hat{x} - x_\mathit{ref})\big)$ to compensate for errors and disturbances while keeping the controller feasible. The saturation limits are detailed in Section~\ref{sec.AboutSaturation}, let's focus now on the linear regime:

The real robot motion
\begin{equation}\label{eq.RealMotion}
   x^{+} = Ax + Bu_\mathit{ref}+BK(x-x_\mathit{ref}) + Be_{u} - BKe_{x}
\end{equation}
deviates away from the reference trajectory because of modelling and estimation errors
\begin{equation}
    e_x = x-\hat{x},\quad e_{x^{\texttt{+}}} = x^+\! -\hat{x}^+\!,\quad Be_u = e_{x^\texttt{+}} - Ae_x,
\end{equation}
resulting on tracking errors $\tilde{x} = x - x_\mathit{ref}$ as
\begin{equation}\label{eq.tr_error_dynamics}
    \tilde{x}^{+} = (A+BK)\tilde{x} + B(e_{u}-Ke_{x}).
\end{equation}

We choose $K$ such that $A+BK$ is strictly stable, so if the errors are bounded to some interval $\mathcal{D}$
\begin{equation}\label{eq.boundedDisturbances}
    e_{u}-Ke_{x} \in \mathcal{D} \subset \mathbb{R},
\end{equation}
the tracking error is also bounded to be in the minimal Robust Positively Invariant (mRPI) set of the disturbed system $\Omega$ \cite{Mrpi}
\begin{equation}
    \tilde{x} \in \Omega,
\end{equation}
which is convex, compact and contains the set $B\mathcal{D}$. Thanks to this bound, we can guarantee that any reference state satisfying \eqref{eq. ReferenceStateConstraint} will result on a feasible real state provided that the set $ \mathcal{X}\ominus \Omega $ is non-empty.

The price to pay for this guarantee is some extra restrictiveness on the reference VRP $ v_\mathit{ref} $ due to the safety margin
\begin{equation}
    \tilde{v}_\mathrm{max} = \max_{\tilde{x}\in\Omega} \; V \tilde{x},
\end{equation}
that corresponds to the maximum VRP tracking error. We can reduce this restrictiveness by minimizing the tracking error bound $\tilde{v}_\mathrm{max}$ with an appropriate choice of the feedback gain $K$ \cite[Chapter 4.5]{Villa2019managing}:
\begin{equation}\label{eq.OptimalGain}
K_\mathit{opt} = \arg\!\min_{\hspace*{-3mm}K} \;\;\;\tilde{v}_\mathrm{max},
\end{equation}
which can be obtained, for example, with a derivative free solver such as \textit{Neilder-Mead}~\cite{Gao_2012_Neilder-Mead}.

\subsection{Saturation limits}\label{sec.AboutSaturation}

We saturate the feedback signal to make sure that the resulting next state is feasible. Considering that the reference motion is feasible by design, we can obtain from \eqref{eq.DynamicalConstraint} that the tracking error must satisfy:
\begin{equation}
    V\tilde{x}^+ \in \mathcal{P}(s) - Vx_\mathit{ref}^+ - n^+.
\end{equation}
Introducing the dynamics \eqref{eq.tr_error_dynamics}, this implies a saturation function for each component $^{xy}$ as
\begin{align}
\mathrm{sat}\big(K\hat{\tilde{x}}\big) \!=
\!\left\{
        \!\!\begin{array}{lll}
            {(K\tilde{x})}_\mathrm{min} & \mathrm{if} &  K\hat{\tilde{x}} \leq {(K\tilde{x})}_\mathrm{min} \\
            K\hat{\tilde{x}} & \mathrm{if}& {(K\tilde{x})}_\mathrm{min} \leq K\hat{\tilde{x}} \leq {(K\tilde{x})}_\mathrm{max} \\
            {(K\tilde{x})}_\mathrm{max} & \mathrm{if} & {(K\tilde{x})}_\mathrm{max} \leq K\hat{\tilde{x}}
        \end{array}
    \right.
\end{align}
with the limits:
\begin{align}
    ({K\tilde{x})}_\mathrm{min} &= \dfrac{p^{ }_\mathrm{min}-n^+ - V\!(A\tilde{x}+x_\mathit{ref}^+)}{V\!B} - e_{\mathit{u},\mathrm{min}},\\
    ({K\tilde{x})}_\mathrm{max} &= \dfrac{p^{ }_\mathrm{max}-n^+ - V\!(A\tilde{x}+x_\mathit{ref}^+)}{V\!B} - e_{\mathit{u},\mathrm{max}},
\end{align}
where $\hat{\tilde{x}}$ stands for the measured tracking error $ \hat{x} - x_\mathit{ref} $ and $V\!B$ is a scalar value.


\section{WHOLE-BODY CONTROL}\label{Sec.WholeBodyControl}

We formulate an Inverse Dynamics (ID) problem to compute optimally all motor torques 
$\tau_{a}$, contact wrenches $ \phi_k $, and accelerations $ \hat{\ddot{q}} $ subject to the rigid robot model \eqref{eq.LagrangeanDynamics} and non-sliding condition \eqref{eq. NonSlidingCondition} with the estimated posture $\hat{q}$.
%
%
As standard, this optimization problem is composed by tasks that define the control goals and constraints \cite{adelprete:jnrh:2016, Noelie_2022_TSID}. 
\subsection{Task descriptions}
In the \textit{Interface} block, we translate the desired centroidal motion $ s_\mathit{ref}^+ $, $ \hat{x}^{+} $ into task references, seeking to avoid conflicts between them in order to reduce the incidence of weight tuning and to obtain contact forces as close as possible to the stabilizer solution:
\begin{itemize}
    \item[] \hspace*{-7mm}\textbf{Center of Mass}: The desired values $c^\mathit{des}$, $\dot{c}^\mathit{des}$, $\ddot{c}^\mathit{des}\in \mathbb{R}^3$ are obtained from the centroidal state $\hat{x}^{+}$.
    
    \item[] \hspace*{-7mm}\textbf{Feet Motion}: Right and left positions $s^\mathit{des}_{R}$, $s^\mathit{des}_{L}\in \mathit{SE}(3)$ as well as their time derivatives, are obtained from splines connecting the feet placements $s_\mathit{ref}$ \cite{ndcurves}.
    
    \item[] \hspace*{-7mm}\textbf{Waist Orientation}: The desired orientation $R^\mathit{des}_{w}\in \mathbb{R}^{3}$, composed by three Euler angles, maintains zero roll and pitch rotations, with the yaw angle as the bisector between right and left feet.
    
    \item[] \hspace*{-7mm}\textbf{Posture}: All joint angles in the robot legs are analytically computed as proposed in \cite[Chapter 2.5]{Kajita2014Introduction} to agree with the desired CoM and feet positions. The full desired posture $q_{j}^\mathit{des}$, $\dot{q}_{j}^\mathit{des}$, $\ddot{q}_{j}^\mathit{des}\in \mathbb{R}^{32}$ is completed with fixed torso and arm references.
    
    \item[] \hspace*{-7mm}\textbf{Angular Momentum}: The desired Angular momentum $L^\mathit{des}$, $\dot{L}^\mathit{des}\in \mathbb{R}^{3}$ is obtained according to the desired configuration $\dot{q}^\mathit{des}$, $\ddot{q}^\mathit{des}\in \mathbb{R}^{38}$
    \begin{align}
        L^\mathit{des} &= G_\textsc{am} \dot{q}^\mathit{des},\\
        \dot{L}^\mathit{des} &= G_\textsc{am} \ddot{q}^\mathit{des} +\dot{G}_\textsc{am} \dot{q}^\mathit{des},
    \end{align}
    using the angular part of the centroidal momentum matrix $G_\textsc{am}$.
    
    \item[] \hspace*{-7mm}\textbf{Force Distribution}: We obtain ground contact wrenches $\phi^\mathit{des}_{R}$, $\phi^\mathit{des}_{L}\in \mathbb{R}^{6}$ from Newton and Euler equations considering the desired values $c^\mathit{des}$, $\ddot{c}^\mathit{des}$, $\dot{L}^\mathit{des}$. During single support stages the desired centroidal motion determine one unique wrench $\phi^\mathit{des}$, but during double support stages, we manage the redundancy with quadratic programming by minimizing each wrench magnitude.
    
\end{itemize}

\subsection{Policy in task spaces}
\noindent Joint motion related tasks use forward-kinematic functions $ \gamma(q) $ and their time derivatives
\begin{align}
    \dot{\gamma} = \dfrac{d\gamma}{d\hat{q}}\hat{\dot{q}} = J_\mathit{task}\hat{\dot{q}},\\
    \ddot{\gamma} = J_\mathit{task}\hat{\ddot{q}} + \dot{J}_\mathit{task}{\hat{\dot{q}}},
\end{align}
with task-specific Jacobians $J_\mathit{task}$, to approach desired values $ \gamma^\mathit{des}$, $\dot{\gamma}^\mathit{des}$, $\ddot{\gamma}^\mathit{des} $. Each task can be set as a cost function, for the ID to minimize the square norm
\begin{equation}
    V_\mathit{task} = \Vert \ddot{\gamma} - \pi_\mathit{task}\Vert^2,
\end{equation}
or as a constraint, for the ID to impose the value
\begin{equation}
    \ddot{\gamma} = \pi_\mathit{task},
\end{equation}
both using a feedback law $\pi_\mathit{task}$, with gains $ K_\textsc{p}^\mathit{task} $, $ K_\textsc{d}^\mathit{task} $, normally used to produce task consistent accelerations \cite{Sherikov2014Whole-BalanceConst}
\begin{equation}
    \pi_\mathit{task} = K^\mathit{task}_\textsc{p}(\gamma-\gamma^\mathit{des})+ K^\mathit{task}_\textsc{d}(\dot{\gamma}-\dot{\gamma}^\mathit{des}) + \ddot{\gamma}^\mathit{des}.
\end{equation}

We formulate similarly an angular momentum task as a cost function:
\begin{align}
    V_\textsc{am} &= \Vert G_\textsc{am}\hat{\ddot{q}}+\dot{G}_\textsc{am}\hat{\dot{q}} - \pi_\textsc{am}\Vert^2,\\
    \pi_\textsc{am} &= \dot{L}^\mathit{des} \!+\! K_\textsc{p}^\textsc{am}(G_{\textsc{am}}\hat{\dot{q}}-L^\mathit{des}),  
\end{align}
using the angular part $G_\textsc{am}$ of the centroidal momentum matrix \cite{Orin2008Centroidal}.

Contact force related tasks are formulated to minimize square norms of the form 
\begin{equation}
    V_\mathit{task} = \Vert D_\mathit{task}\phi_k - D_\mathit{task}\phi_k^\mathit{des}\Vert^2
\end{equation}
for the wrench $ \phi_k $ on the $k$-th contact, with an appropriately chosen matrix $D_\mathit{task}$.

\section{EXPERIMENTS}\label{Sec.Experiments}


\subsection{Experimental Setup}

Our experiments are performed with the torque controlled robot Talos, using its internal computer that has an Intel(R) Core(TM) i7-3612QE (4 cores, 2.10GHz) and 16Gb of RAM. The robot also includes 6D force/torque sensors at the ankles and wrists.

In all the experiments described below, we use the full control diagram displayed in Fig.~\ref{Fig.ControlDiagram}. Though, we may disable the MPC (using a precomputed reference motion) or the bending estimator when explicitly said. In our implementation, the MPC runs at $ 5 $~Hz while the rest (centroidal stabilizer, estimators, interface and  inverse dynamics) runs at the higher frequency of $500$~Hz.


We have chosen the feedback gains 
\begin{equation}
    K = \begin{bmatrix} \texttt{-}9894 & \texttt{-}4189 & \texttt{-}496 \end{bmatrix},
\end{equation}
obtained from \eqref{eq.OptimalGain}, according to the robot disturbances.

We set the MPC with the safety margins $v_\mathrm{max}^x = 2.5$~cm and $v_\mathrm{max}^y = 1.5$~cm in the support polygon, which guarantees feasibility with disturbances up to $e_u^x + Ke_x^x = 5370$~m/s$^3$ and $e_u^y + Ke_x^y = 3222$~m/s$^3$ following \cite[Chapter 4]{Villa2019managing}.

\subsection{Stiffness Identification}\label{Sec.stiffnessIdentification}

We set up the robot to keep balance statically on one foot over a horizontal ground. In such conditions, the acceleration \eqref{eq.LinearizedCentroidalDynamics} and bias \eqref{eq.Non-linearEffects} vanishes $\ddot{c} = 0$, $n = 0$, so that 
\begin{equation}\label{eq.EquilibriumCondition}
    p^{xy}=v^{xy}=c^{xy}
\end{equation}
follows from \eqref{eq.LinearizedCentroidalDynamics} and \eqref{eq.VRP_and_COP}.

We estimate the CoM $c^\mathrm{fk}(\hat{q})$ and supporting foot position $s^\mathrm{fk}(\hat{q})$ by forward kinematics \cite{Cotton2009estimation} using the corrected posture~$\hat{q}$ (see Sec. \ref{Sec.CompensationOfDeflections}).

In both estimations, we correct the expected rigid model errors $\Delta_i = R(\theta_i^{y x})l_i\,-\,l_i$, produced by the lever arm $l_i$ on each leg $i\in\{ \mathit{left},\;\mathit{right}\}$, as
\begin{align}
    s_i = s_i^\mathrm{fk}(\hat{q}) + \Delta_i,\qquad c = c^\mathrm{fk}(\hat{q}) + \dfrac{\sum_i \Delta_i m_i}{m},
\end{align}
with $m_i$ and $m$ standing for the leg and whole robot masses respectively.
%
%

\begin{figure}
	\centering
	\includegraphics[trim={0 0.5mm 0 3.5mm},clip,width=0.45\textwidth]{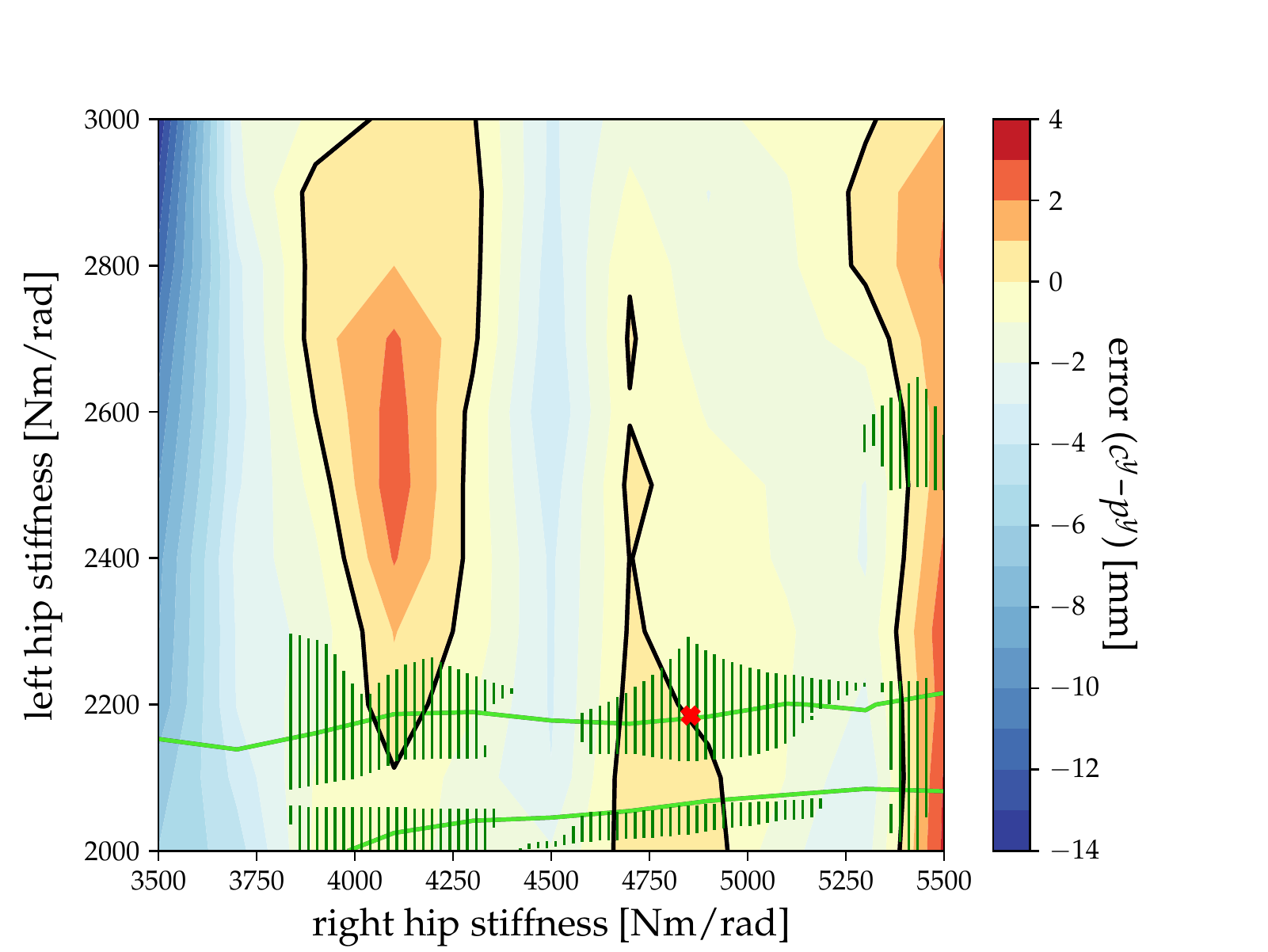}
	\caption{\textbf{Stiffness Identification:} Error contour curves obtained from the right support experiment. The zero error level is highlighted on \textit{black}, and the \textit{light green} curve shows the zero error level obtained from the left support experiment. The identified stiffness belongs to the \text{green} patches, corresponding to zero error on both experiments, within two standard deviations of our experimental measurement.}
	\label{Fig:flex_id}
	\vspace{-3mm}
\end{figure}

We also measure the sole torque $\tau_\mathrm{sole}^{xy}$ and ground normal force $f_\mathrm{sole}^z$ using the 6D force/torque sensor on the robot ankle to obtain the CoP as:
\begin{equation}
    p^{xy} = s_i^{xy} + S\,\frac{\tau_\mathrm{sole}^{xy}}{f_\mathrm{sole}^z},
\end{equation}
where $ S=\big[\begin{smallmatrix}\,0\, & \texttt{-}1\\1\, &0 \end{smallmatrix}\big]$. 
With the robot supported on one foot, we vary the assumed stiffness of both hips (and therefore the estimated posture $\hat{q}$) on a grid corresponding to the axes of the Fig.~\ref{Fig:flex_id}. The measured error $c^{y}\texttt{-}p^{y}$ is reported by the level curves throughout the grid of tested stiffness.

We carry out this experiment twice, standing first on the left support and then on the right support. The correct stiffness values must satisfy the condition \eqref{eq.EquilibriumCondition} ($c^{xy}\!~-~\!p^{xy}\!~=~\!0$) on both experiments.

We have identified a stiffness of $k_f = 4900 \pm 200$~Nm/rad on the right hip and $k_f = 2180 \pm 70$~Nm/rad on the left hip, marked with a red cross. We approximate the values of damping to be conservatively small as $d_f = 2\sqrt{k_f}$.

\subsection{Quasi-Static Walking}
\begin{figure}[t!]
	\centering
	\includegraphics[trim={0 1mm 0 3mm},clip,width=0.45\textwidth]{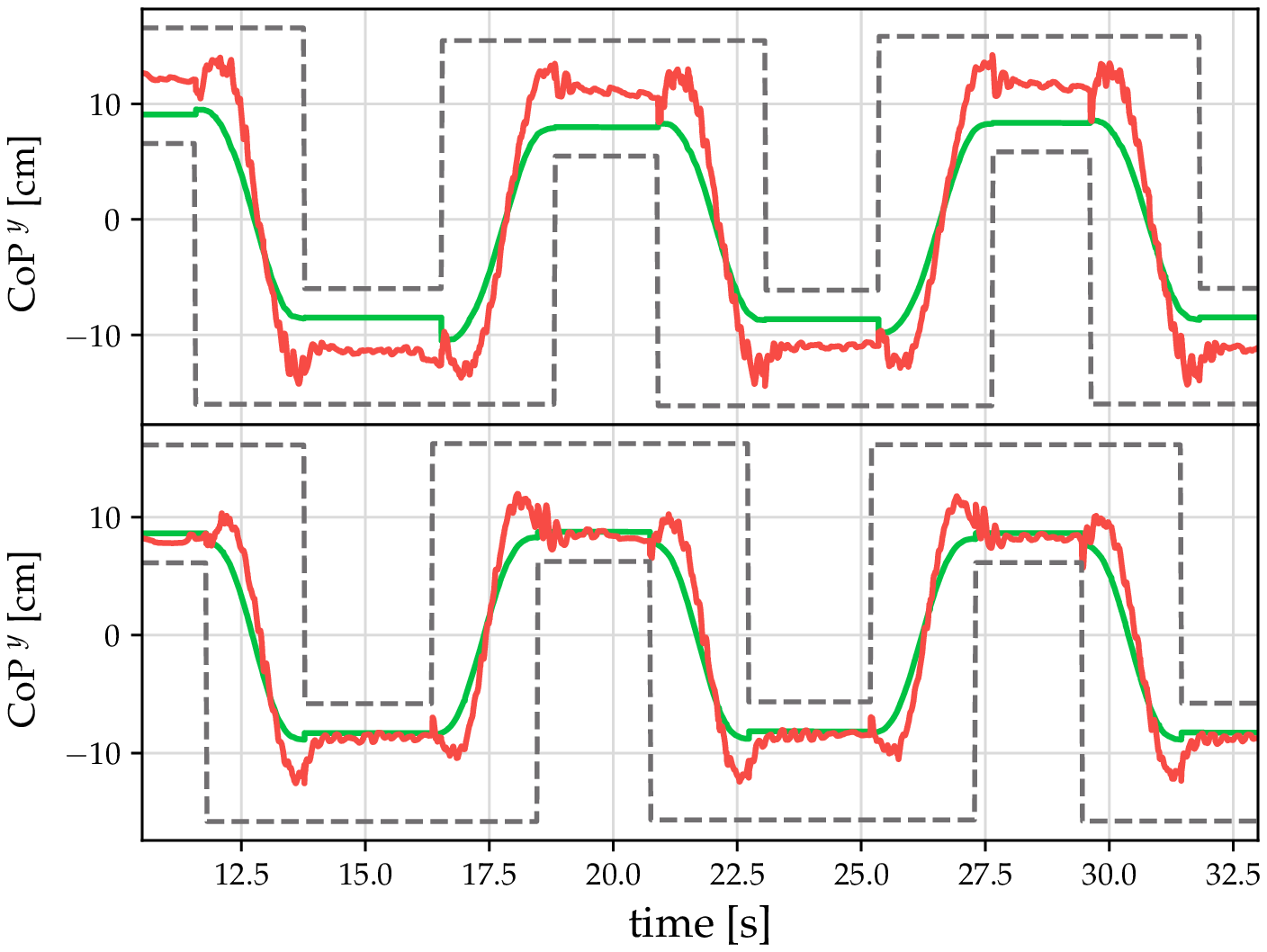}
	\caption{\textbf{Quasi-static / walk:} Reference (\textit{green}) and real (\textit{red}) CoP when the robot posture is estimated without taking hip flexibility into account (\textit{up}) and considering an estimation of deflections as discussed above (\textit{down}). The robot support is shown with \textit{gray} dashed lines.} \label{Fig.quasisteticWalk}
    \vfill 
	\centering
	\includegraphics[trim={0 1mm 0 3mm},clip,width=0.45\textwidth]{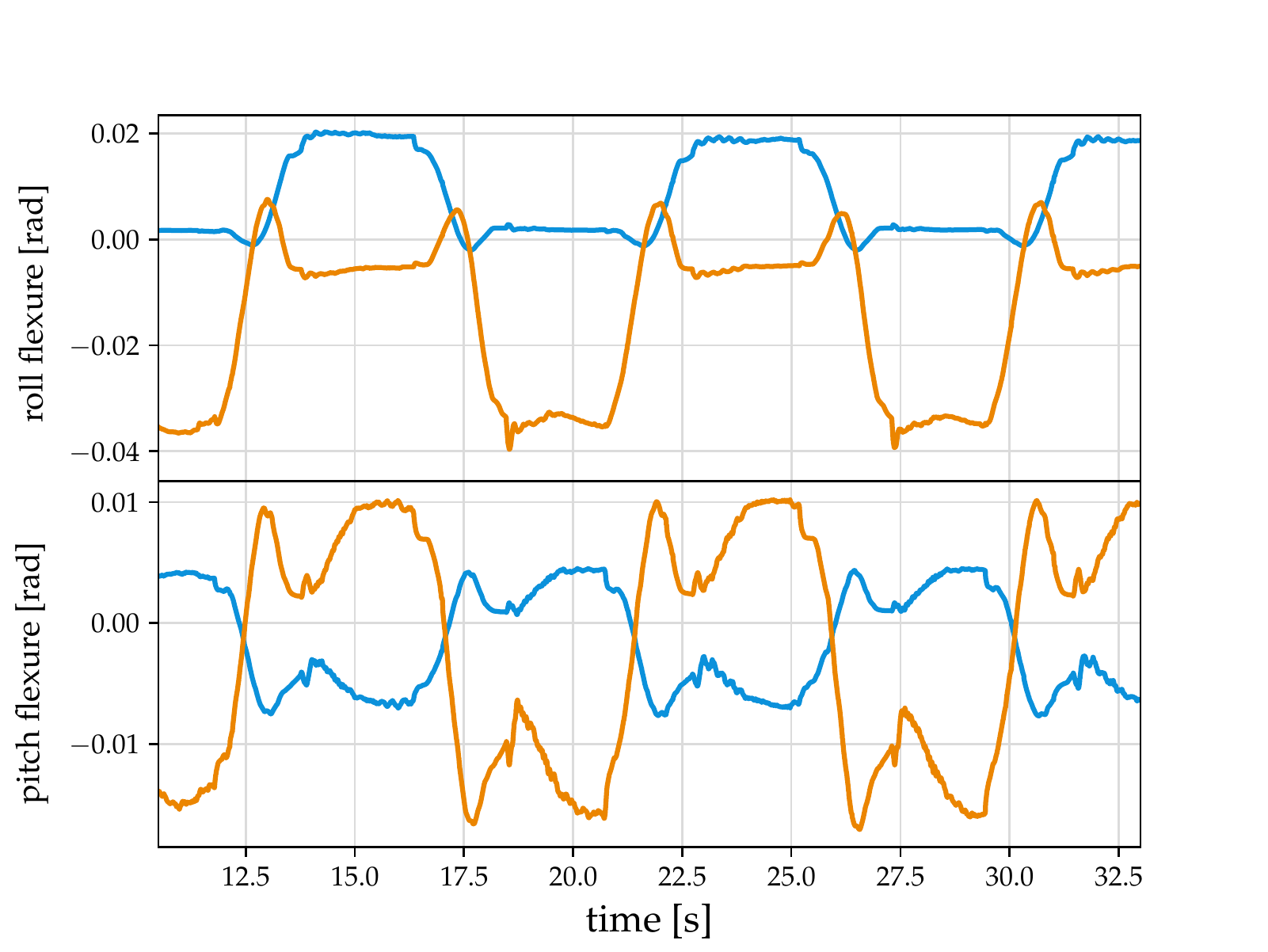}
	\caption{\textbf{Quasi-static / Deflections:} Roll (\textit{up}) and pitch (\textit{down}) deflections, estimated during the experiment with the bending estimator activated. Left hip deflections are shown in \textit{orange} and right hip deflections are shown in \textit{blue}.}
	\label{Fig:Quasi-Static-Deflections}
\end{figure}

    In a quasi-static walk, the robot adopt a sequence of statically stable postures in which the CoM is always maintained over the support polygon and the CoM acceleration is assumed to be zero. This kind of motion has been achieved on the robot without estimation of deflections by just setting the reference CoP trajectories with an offset towards the interior bound of feet, this can be observed on Fig.~\ref{Fig.quasisteticWalk}. 
    
    We compare the robot behavior when tracking a precomputed quasi-static reference trajectory with and without bending estimator on Fig.~\ref{Fig.quasisteticWalk}. In both experiments the robot walks forward making steps of $10$~cm. As some CoM acceleration is needed to switch between supporting feet, we can see growing CoP tracking errors $\tilde{p} = p\texttt{-}p_\mathit{ref}$ at switching times for both experiments. We can also see that, during single support periods (when the hip torque is maximal), hip deflections produce and maintain a big CoP tracking error ($\approx 3$~cm) which we have nearly removed thanks to a more precise posture estimation. 
    
    The estimated pitch and roll deflections during the experiment with bending estimator are illustrated on the Fig.~\ref{Fig:Quasi-Static-Deflections}. Left and right deflections have different magnitudes due to their different values of stiffness. As Talos legs lengths are of $90$~cm, if not identified, such deflections may produce feet misplacements of approximately $3$~cm which correspond to the CoP tracking error observed without the estimator.

\begin{figure}
	\centering
	\includegraphics[trim={0 1mm 0 3mm},clip,width=0.45\textwidth]{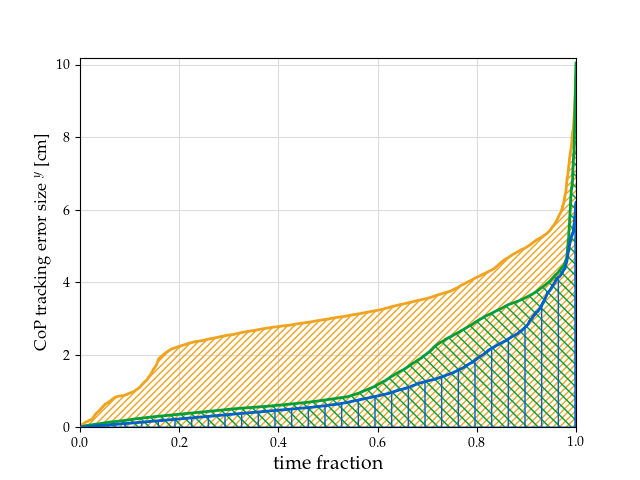}
	\caption{\textbf{Tracking errors:} The CoP tracking error is maintained in the shadowed area (below the curve) during the fraction of experiment time indicated on the horizontal axis. The cases of quasi-static walks with and without bending estimator are respectively shown in \textit{green} and \textit{orange}. The dynamic walk case with bending estimator is shown in \textit{blue}.}
	\label{Fig:Quasi-Static-TrackingError}
\end{figure}

    The Fig.~\ref{Fig:Quasi-Static-TrackingError} indicates the fraction of time during which the tracking error is maintained below each size. It makes clear that the bending estimator reduces the tracking error along the entire experiment, maintaining it below $1$~cm during $55$\% of the time, while the experiment without the estimator has tracking errors above $3$~cm during $50$\% of the time. The tracking error peak of $10$~cm happens in both cases when the robot start walking.

    We can conclude from these experiments that, by taking hip deflections into account, we make a better estimation of the contact points improving, therefore, the control of forces.
    
\subsection{Dynamic Walking}

In a dynamic walk, the robot is allowed to move its CoM outside the support polygon between steps by adopting dynamically-stable postures.
We have only achieved this kind of motion by taking hip deflections into account. 

\begin{figure}[t!]
	\centering
	    \includegraphics[trim={0 1mm 0 3mm},clip,width=0.45\textwidth]{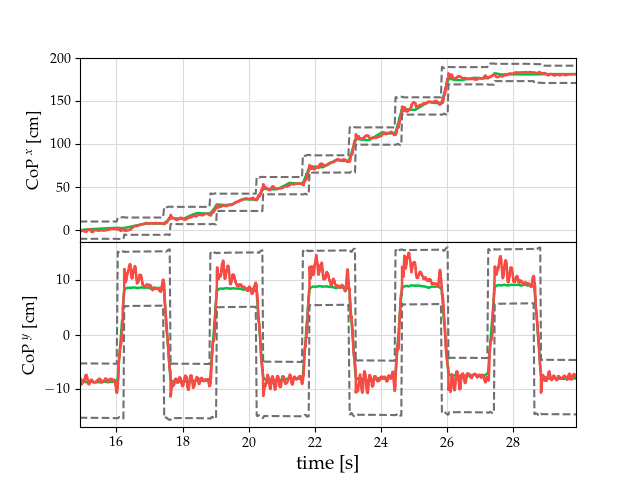}
    	\caption{\textbf{Dynamic walk:} Reference (\textit{green}) and real (\textit{red}) CoP obtained in both coordinates $^x$ (\textit{up}) and $^y$ (\textit{down}) when the forward motion starts. The robot support is shown with \textit{gray} dashed lines. }
	    \label{Fig:Dynamic_walk}
	\vfill    
	\centering
	    \includegraphics[trim={0 1mm 0 3mm},clip,width=0.45\textwidth]{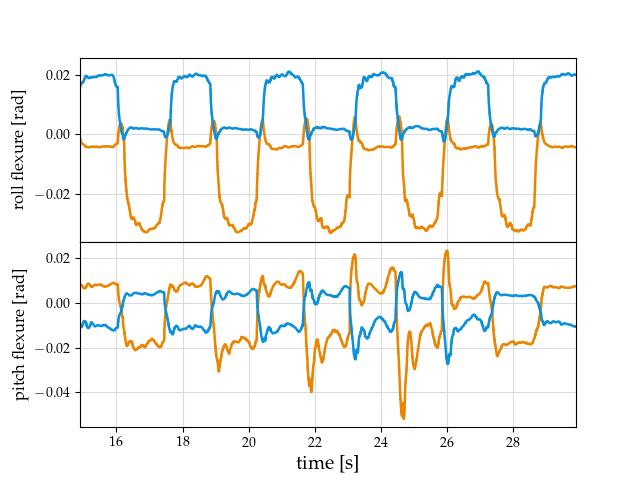}
    	\caption{\textbf{Dynamic / Deflections:} Roll (\textit{up}) and pitch (\textit{down}) deflections, recorded during the dynamic walking when the forward motion starts. Left hip deflections are shown in \textit{orange} and right hip deflections are shown in \textit{blue}.}
    	\label{Fig:Dynamic_Flexure}
\end{figure}

In order to show the behavior of hip deflections with different forward velocities, we perform a walking experiment where the robot starts walking in place, and then, it makes $8$ growing steps forward of up to $35$~cm taking $1.4$~s per step
($1.2$~s in single support and $0.2$~s in double support)%
. This experiment is recorded in \href{https://gepettoweb.laas.fr/articles/talos_centroidal_mpc_torque_control.html}{video}\footref{video} and its centroidal data is shown in Fig.~\ref{Fig:Dynamic_walk}. 

At each step, big CoP$^y$ oscillations appear and are quickly stabilized. We think these oscillations may be produced by our bad estimation of the hip damping coefficients or a too high cutting frequency in the filter \eqref{eq. Deflections}. Nevertheless, the whole robot motion is smooth and fluid as illustrated in the \href{https://gepettoweb.laas.fr/articles/talos_centroidal_mpc_torque_control.html}{video}\footref{video}.

Fig.~\ref{Fig:Quasi-Static-TrackingError} shows that the tracking error during this experiment is even smaller than in the quasi-static case, staying with errors below $1$~cm during $65$\% of the experiment.

Hip deflections, shown in Fig.~\ref{Fig:Dynamic_Flexure}, are similar to those of quasi-static walk when walking in place, but when walking forward, the pitch torque required to move the legs grows with the step length and swinging velocity producing each time bigger and faster deflections, which result in also growing CoP$^x$ tracking errors.

%


The locomotion obtained in this and other experiments, shown in the \href{https://gepettoweb.laas.fr/articles/talos_centroidal_mpc_torque_control.html}{video}\footref{video}, is smooth, stable and reaches velocities up to $25$~cm/s, faster than the outstanding results obtained in \cite{MesesanEGOA19} for the robot Toro ($15$~cm/s) with similar settings. We have yet to explore the use of edge contacts for even faster walks, as done in \cite{MesesanEGOA19} to reach $37$~cm/s.


\section{CONCLUSIONS}\label{Sec.Conclu}

In this work, we have proposed a full control scheme to produce fast and dynamic locomotion in torque control with heavy robots that present link flexibility.

We have paid special care to reduce all model and control errors within a robust approach by taking link deflections into account to estimate the robot posture, taking the full centroidal dynamics into account to generate reference trajectories, and making all whole-body references internally consistent. Then, we compensate for all residual errors using state feedback, with feasibility guaranteed for bounded errors.

The most evident outcome of all this modelling and control effort is the level of locomotion it enabled in Talos (\href{ https://gepettoweb.laas.fr/articles/talos_centroidal_mpc_torque_control.html}{video}\footref{video}), with walking velocities of up to $25$~cm/s, among the best performances reached so far with electric robots in torque control. But, moreover, we have demonstrated that link flexibility has important effects on heavy robots such as Talos, motivating more specific control approaches \cite{Giulio2022_TalosFlex}.

Finally, we also provide a simple experimental method to identify stiffness in Talos' hips that does not require additional equipment.

\bibliography{cleanBib.bib}
\bibliographystyle{ieeetr}

\end{document}